\documentclass[sigconf, screen]{acmart}

\AtBeginDocument{%
  \providecommand\BibTeX{{%
    \normalfont B\kern-0.5em{\scshape i\kern-0.25em b}\kern-0.8em\TeX}}}

\usepackage{colortbl}
\usepackage{enumitem}
\usepackage{multirow}

\definecolor{azure}{rgb}{0.663, 0.780, 0.941}
\setlist{nosep}
\newcommand{\tabitem}{\textbullet~~}

\copyrightyear{2024}
\acmYear{2024}
\setcopyright{rightsretained}
\acmConference[IVA '24]{ACM International Conference on Intelligent Virtual Agents}{September 16--19, 2024}{GLASGOW, United Kingdom}
\acmBooktitle{ACM International Conference on Intelligent Virtual Agents (IVA '24), September 16--19, 2024, GLASGOW, United Kingdom}\acmDOI{10.1145/3652988.3673929}
\acmISBN{979-8-4007-0625-7/24/09}

\newcommand{\repourl}{https://github.com/hyesunyun/va-item-variants}

\begin{document}

\title[Keeping Users Engaged During Repeated Interviews by a Virtual Agent]{Keeping Users Engaged During Repeated Interviews by a Virtual Agent: Using Large Language Models to Reliably Diversify Questions}

\author{Hye Sun Yun}
\email{yun.hy@northeastern.edu}
\affiliation{%
  \institution{Northeastern University}
  \city{Boston}
  \state{MA}
  \country{USA}
}

\author{Mehdi Arjmand}
\email{arjmand.me@northeastern.edu}
\affiliation{%
  \institution{Northeastern University}
  \city{Boston}
  \state{MA}
  \country{USA}
}

\author{Phillip Sherlock}
\email{phillip.sherlock@ufl.edu}
\affiliation{%
  \institution{University of Florida}
  \city{Gainesville}
  \state{FL}
  \country{USA}
}

\author{Michael K. Paasche-Orlow}
\email{mpo@tufts.edu}
\affiliation{%
  \institution{Tufts University}
  \city{Boston}
  \state{MA}
  \country{USA}
}

\author{James W. Griffith}
\email{jamesgriffith@uchicago.edu}
\affiliation{%
  \institution{University of Chicago}
  \city{Chicago}
  \state{IL}
  \country{USA}
}

\author{Timothy Bickmore}
\email{t.bickmore@northeastern.edu}
\affiliation{%
  \institution{Northeastern University}
  \city{Boston}
  \state{MA}
  \country{USA}
}

\renewcommand{\shortauthors}{Yun, et al.}

\begin{abstract}
  Standardized, validated questionnaires are vital tools in research and healthcare, offering dependable self-report data. Prior work has revealed that virtual agent-administered questionnaires are almost equivalent to self-administered ones in an electronic form. Despite being an engaging method, repeated use of virtual agent-administered questionnaires in longitudinal or pre-post studies can induce respondent fatigue, impacting data quality via response biases and decreased response rates. We propose using large language models (LLMs) to generate diverse questionnaire versions while retaining good psychometric properties. In a longitudinal study, participants interacted with our agent system and responded daily for two weeks to one of the following questionnaires: a standardized depression questionnaire, question variants generated by LLMs, or question variants accompanied by LLM-generated small talk. The responses were compared to a validated depression questionnaire. Psychometric testing revealed consistent covariation between the external criterion and focal measure administered across the three conditions, demonstrating the reliability and validity of the LLM-generated variants. Participants found that the variants were significantly less repetitive than repeated administrations of the same standardized questionnaire. Our findings highlight the potential of LLM-generated variants to invigorate agent-administered questionnaires and foster engagement and interest, without compromising their validity.
\end{abstract}



\begin{teaserfigure}
  \includegraphics[width=\textwidth]{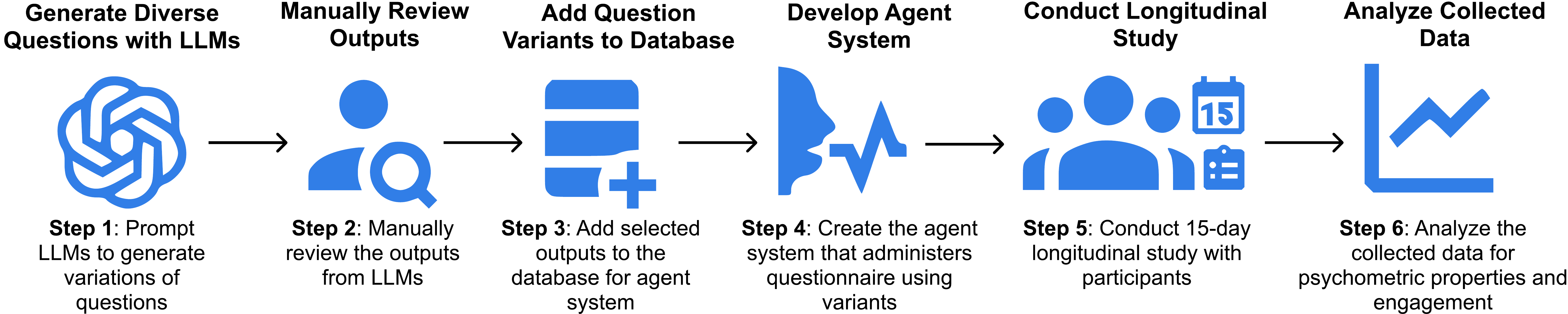}
  \caption{A workflow diagram of the longitudinal validation study which evaluated the validity, reliability, and user engagement of utilizing large language model-generated variants of a standardized depression questionnaire.}
  \Description{A workflow diagram of the study is presented from left to right. Six components: Generate Diverse Questions with LLMs, Manually Review Outputs, Add Item Variants to Database, Develop Agent System, Conduct Longitudinal Study, and Analyze Collected Data. Step 1 is the OpenAI logo indicating the prompting of LLMs to generate diverse questions. Step 2 is a person with a magnifying glass indicating manually reviewing the outputs from the LLMs. Step 3 is a database logo with a plus sign to indicate adding manually selected outputs to the database for the virtual agent system. Step 4 is a talking person indicating the creation of the virtual agent system for administering the questionnaire using the variants. Step 5 is groups of people with a calendar with a 15 on it and a survey logo indicating the 15-day longitudinal study. Lastly, step 6 consists of a graph logo which shows analyzing collected data for psychometric properties and engagement.}
  \label{fig:diagram_teaser}
\end{teaserfigure}

\maketitle

\section{Introduction}
\label{sec:intro}

Self-report questionnaires are a type of self-report method that includes a set of questions in a highly structured, standardized written form. Validated questionnaires are widely used in research and healthcare as an assessment strategy as they offer dependable self-report data. Prior work has revealed that human and virtual agent-administered questionnaires are nearly equivalent to self-administered questionnaires in the electronic form \cite{jaiswal2019virtual,bickmore2020substance}. These studies have shown the feasibility and reliability of using virtual agents (VAs) to administer questionnaires simulating interviews for a single session. However, many repeated-measures evaluation studies and longitudinal interventions require the same self-report questionnaire to be administered to the same individual multiple times.
In healthcare, patient-reported outcomes (PROs) are used to obtain self-reports of a patient's condition at home, typically involving the repeated administration of surveys to capture symptoms or quality of life \cite{Fayanju16}.

However, response rates to repeated surveys tend to decline over time, as respondents become fatigued by repeatedly filling out the same questionnaires \cite{porter2004multiple,Min14,Dean16}. Even in healthcare, where PROs can be used as the basis for treatment decisions, longitudinal survey completion rates can be as low as 48\% \cite{Min14,Dean16,Huynh21}. Dwindling response rates can lead to a nonresponse measurement bias \cite{Groves08} and limit the ability to evaluate important changes over time.

Several strategies have been proposed to increase repeated measure response rates, including incentives \cite{VanGeest07}, more frequent contact and engagement with respondents \cite{Cleary15,Prink19}, and providing survey responses back to the individuals being surveyed \cite{Vemuru23}. In addition, a variety of approaches have been studied to increase the usage rates for repeated interactions with VAs, including the use of syntactic and visual variability in the interface \cite{bickmore2010maintaining} and humor \cite{Olafsson20}. Other strategies for automated systems include reminders \cite{Hernandez20} and social support and reinforcement \cite{Smith08}.

In this work, we explore two strategies to increase response rates to a PRO administered daily for two weeks by a virtual agent (VA) that simulates a face-to-face interview with a healthcare professional. The first strategy involved using survey questions that vary in every administration so that the survey administrations sound different. We went beyond straightforward syntactic variation of questions to variants generated by large language models (LLMs) to capture the latent construct we are interested in. Syntactic variations primarily entail the reordering of words within a sentence, whereas LLM-generated variations we employed had slightly different words or phrases that convey comparable meanings. Second, we explore the use of small talk, humor, and empathy generated by LLMs to make daily interactions with the VA more conversational, entertaining, and engaging.

Many questionnaires, including most PROs, have been validated using laborious methods involving testing with dozens, if not thousands, of respondents to establish reliability and validity \cite{Furr21}. An important question raised when validated questionnaires are modified is whether the new derivative versions retain the reliability and validity of the original form. We report the results of a longitudinal study involving a validated PRO for depression, in which participants engaged with our virtual agent system daily for two weeks. Participants were randomized to either a repeated, standardized depression questionnaire or one of the two interventions with LLM-generated questionnaire variants. All participants completed an additional standardized, validated depression questionnaire which was a criterion for comparison. Our hypotheses are:

\begin{itemize}
\item \textbf{H1}: VA administration of LLM-generated questionnaire variants will retain similar validity and reliability to the VA administration of the original questionnaire.
\item \textbf{H2}: Questionnaires delivered in a different form using LLM-generated variants daily will be more engaging for participants, based on the number of questionnaires completed and feedback from participants.
\item \textbf{H3}: Questionnaires delivered with LLM-generated conversational small talk, humor, and empathy will be more engaging compared to those delivered as strictly question-and-response interviews by a VA.
\end{itemize} 

Our primary contributions include the introduction of an innovative approach using LLMs to reliably generate diverse versions of validated questionnaires for a VA system. Furthermore, we demonstrate the feasibility of employing these questionnaire variants to enhance user engagement and mitigate repetitiveness.

\section{Related Work}
\label{sec:related_work}

Our work draws on previous research on alternative delivery methods for questionnaires, user engagement methods for longitudinal research, and applications of LLMs to agents and surveys. 
Traditionally, paper or mail surveys have been used to administer questionnaires. However, web-based or online surveys have been more frequently employed, as response rates to paper surveys have declined over time \cite{de2002trends,shannon2002comparison,roster2004comparison,evans2005value,link2005alternative}. However, web-based surveys are not immune to dwindling response rates \cite{vehovar2002nonresponse,manfreda2008web}. Particularly for long or repetitive administrations of surveys in research, respondents can experience fatigue, which can lower the completion rates and data quality of the responses \cite{porter2004multiple,bowling2005mode,sinickas2007finding,le2021national}. Increasing the quality of self-report data and completion rates through surveys remains an important challenge for researchers to overcome.

\subsection{Computers \& Agents for Quality Self-Report}
\label{sec:comps_agents_self_report}

Several past studies have shown that using computers to administer surveys and interviews can lead to greater self-disclosure, especially for sensitive information in the context of healthcare, as the pressure to respond in socially desirable ways is reduced \cite{weisband1996self, turner1998adolescent, kissinger1999application, newman2002differential, lucas2014s, devault2014simsensei}. Several studies have expanded this approach by incorporating VAs, as research has indicated that using conversational interviews for surveys can effectively reduce errors. \cite{schober1997does}. 
In health-screening interviews with VAs, \citeauthor{lucas2014s} \cite{lucas2014s} discovered that individuals who perceived the VA as automated showed reduced fear of self-disclosure and expressed sadness more intensely than those who perceived the VA as human-operated.
A similar study by \citeauthor{schuetzler2018influence} \cite{schuetzler2018influence} demonstrated that people disclose more about their sensitive behavior to a conversational agent than to a human; however, people disclose less when the conversational agent appears to understand. In addition, \citeauthor{kocielnik2021designing} \cite{kocielnik2021designing} used a chatbot called HarborBot to test a conversational approach for social needs screening in emergency departments and compared it with a traditional survey tool. The results revealed that the conversational approach was perceived to be more engaging, caring, understandable, and accessible among low health literacy users than the traditional approach.

Prior work has demonstrated that medical questionnaires or PROs administered by VAs are valid and statistically equivalent to human or self-administered questionnaires \cite{auriacombe2018development, bickmore2020substance}. For example, \citeauthor{jaiswal2019virtual} \cite{jaiswal2019virtual} conducted two sets of studies using mental health questionnaires: one comparing VA administration to standard self-administration, and the second comparing VA administration to an actual human.
The results showed that the questionnaires administered by the VAs were statistically equivalent to human or self-administered questionnaires. Additionally, \citeauthor{mancone2023use} \cite{mancone2023use} showed that voice assistants, such as Alexa, can be used to administer psychological assessment questionnaires as a powerful way to capture attention and engage users emotionally without compromising validity.

With LLMs becoming increasingly capable and prevalent, researchers have begun investigating how LLMs can be employed for self-reporting and survey research. \citeauthor{jansen2023employing} \cite{jansen2023employing} highlighted how LLMs can help overcome some of the challenges in survey research by generating responses to survey items or question-wording. Despite this promising direction, the authors warned about the risks of harmful and inaccurate outputs when using LLMs. Similarly, \citeauthor{kjell2023beyond} \cite{kjell2023beyond} provided a narrative review of how LLMs can potentially be used for psychological assessments using natural language instead of rating scales. Furthermore, one study employed GPT-3 to power a chatbot to collect self-reported data, such as food intake, exercise, sleep, and work productivity \cite{wei2023leveraging}. The authors found that LLMs also provided the ability to maintain context, state tracking, and provide off-topic suggestions.

\subsection{Maintaining Longitudinal Self-Report with Agents}
\label{sec:maintaining_longitudinal_self_report}

Longitudinal studies that require multiple self-reports often have low completion rates \cite{anhoj2004quantitative}. Although VAs increase engagement when administering questionnaires, they still suffer from user disengagement. The length of the first interaction with a VA has been shown to be the primary predictor of the number of healthcare questionnaires completed by a participant \cite{vardoulakis2013social}. The findings showed that longer first interactions can result in fewer completed questionnaires.
Prior work on maintaining engagement in long-term health interventions with VAs by \citeauthor{bickmore2010maintaining} \cite{bickmore2010maintaining} shows that increased variability in agent behavior and giving the agent a human backstory can also lead to increased engagement. 

\subsection{LLMs for Agents}
\label{sec:llms_va_systems}

Recently, several studies have explored the use of LLMs in agent design and implementation. \citeauthor{antunes2023prompting} \cite{antunes2023prompting} prompted LLMs to assist in creating scenarios for socially intelligent agents often used in education and entertainment. They created a pipeline to generate an agent's beliefs, desires, intentions, plans of action, and emotions. Furthermore, other studies have examined how LLMs can generate dialogue utterances for embodied conversational agents such as social robots \cite{sevilla2023using, hanschmann2023saleshat} to mitigate boredom and increase engagement. \citeauthor{olafsson2023accomodating} \cite{olafsson2023accomodating} explored how LLMs can be used as part of VAs for health applications by incorporating GPT-2 in a hybrid dialog system for a virtual alcohol misuse counselor. GPT-2 generated responses were combined with a rule-based approach to transition through structured counseling sessions. Similarly, our study takes a rule-based approach but incorporates diverse messages generated by LLMs. Due to the sensitive nature of mental health questionnaires, we decided on a human-in-the-loop approach due to LLMs' potential harms in the healthcare context \cite{bickmore2018patient,yun2023appraising, haltaufderheide2024ethics}

\subsection{LLMs for Generating Diverse Text}
\label{sec:llms_diverse_content}

In addition to utilizing LLMs for agents, LLMs have been used to generate diverse texts or paraphrases \cite{yu2023large}. \citeauthor{cegin2023chatgpt} \cite{cegin2023chatgpt} conducted a study comparing the quality of crowd-sourced and LLM-generated paraphrases for their diversity and robustness in intent classification. The authors found that ChatGPT is a viable alternative to human paraphrasing. Furthermore, one study showed that while GPT-4 might not necessarily outperform humans in generating diverse motivational messages, it took only 6 seconds to generate one message compared to an average of 73 seconds for humans \cite{cox2023prompting}. Although LLMs may not always provide the most diverse generated text, they are significantly faster and more grammatically correct than humans. To mitigate the challenges of LLMs, \citeauthor{pehlivanouglu2023enhancing} \cite{pehlivanouglu2023enhancing} demonstrated how prompt engineering can enhance lexical diversity, phrasal variations, fluency, relevance, and syntactical differences while preserving the original meaning.

\begin{figure}[t]
    \includegraphics[width=0.9\linewidth]{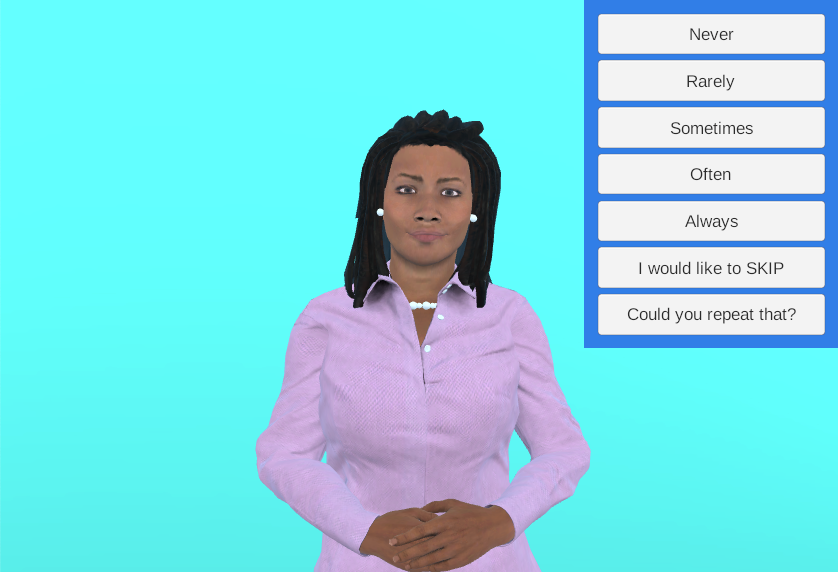}
    \caption{A screenshot of the agent waiting for the user to respond after asking a depression questionnaire question. The dialogue response options are displayed at the top right corner of the screen.}
    \Description{A screenshot of the virtual agent system. An animated character is presented as a woman who stands in the middle of the screen. A rectangular box on the top right is displayed with 7 rectangular buttons with the following labels: Never, Rarely, Sometimes, Often, Always, I would like to SKIP, Could you repeat that?}
    \label{fig:agent}
\end{figure}

\section{System Design}
\label{sec:system_design}

\begin{figure*}[t]
    \includegraphics[width=\textwidth]{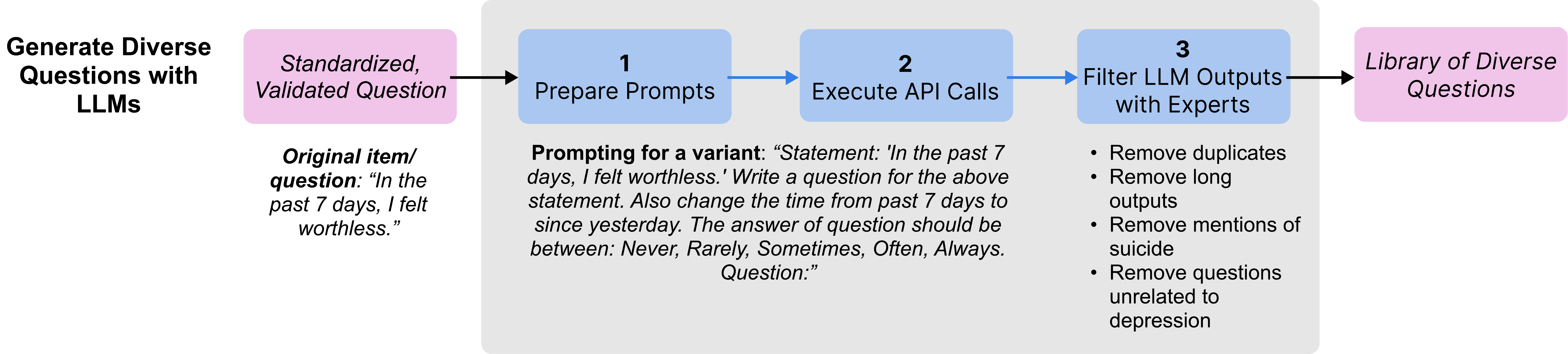}
    \caption{A workflow diagram of how LLMs were used to generate diverse questions. A simple example is provided.}
    \Description{A workflow chart is presented describing how LLMs were used. Five components are sequentially connected in order: Standardized Validated Question, Prepare Prompts, Execute API Calls, Filter LLM Outputs with Experts, and Library of Diverse Questions.}
    \label{fig:process}
\end{figure*}

To evaluate our study hypotheses, we created a VA system deployed over the web for participants to interact daily 
(\autoref{fig:agent}). Our agent is a 3D animated character that converses with users using synthetic speech, conversational behavior, and multiple-choice menu inputs for user responses. The agent's synchronized nonverbal conversational behavior, such as hand gestures, head nods, eyebrow raises, and posture shifts, was automatically generated using the Behavior Expression Animation Toolkit \cite{cassell2004beat}. Agent utterances were generated using template-based text generation. The agent's dialogue is driven by a hierarchical task network-based dialogue engine. The VA system was implemented using the Unity3D game engine and CereProc speech synthesizer.

Our agent, Marie, interacts with participants daily by verbally administering an eight-question questionnaire in dialogue. For our prototype, we focused on one self-report PRO questionnaire using the eight-item PROMIS\textsuperscript{\textregistered} short form depression questionnaire (version 8a) \cite{cella2010initial}. This questionnaire was developed to assess a respondent's level of emotional distress caused by depressed mood where each statement is rated on a five-point scale from 1 being ``Never'' to 5 being ``Always.''

\begin{table*}[t]
  \caption{Depression Questionnaire with Sample Item Variants Generated by LLMs. This table provides the wording variations of the eight-item PROMIS\textsuperscript{\textregistered} short form depression questionnaire \cite{cella2010initial}. All items are rated on a five-point scale from 1=``Never'' to 5=``Always''. \textit{\# of Variants} refers to the number of variants we used for the longitudinal evaluation study.}
  \label{tab:variants}
  \small
  \begin{tabular}{c|m{.13\textwidth}|m{.16\textwidth}|m{.50\textwidth}c}
    \toprule
    \textbf{ID} & \textbf{Original} & \textbf{Control} & \textbf{Sample Variants} & \textbf{\# of Variants}\\
    \midrule
    \rowcolor{gray!25}1 & In the past 7 days, I felt worthless. & In the past 7 days, how often have you felt worthless? & Since we last spoke, have you ever felt like you were a burden to others?; Have you felt like you were not good enough recently?; Since the last time we talked, have you felt like you're not important to anyone? & 8\\
    2 & In the past 7 days, I felt helpless. & In the past 7 days, how often have you felt helpless? & How often have you felt like you were unable to control a situation in the past day?; How often do you feel like you're stuck in a cycle of negativity when faced with challenges?; Have you felt powerless or helpless when dealing with a problem in the past day? How often? & 7\\
    \rowcolor{gray!25}3 & In the past 7 days, I felt depressed. & In the past 7 days, how often have you felt depressed? & Have you been feeling like you can't escape negative thoughts or feelings?; How often have you been feeling empty or numb?; Have you been experiencing changes in your sleep patterns? & 8\\
    4 & In the past 7 days, I felt hopeless. & In the past 7 days, how often have you felt hopeless? & How frequently have you felt like you're drowning in negativity since the last time we talked?; Have you ever felt like everything is pointless, even if things are going well? If so, how often?; Have you felt like you're stuck in a rut or in a situation that's beyond your control? If so, how often? & 7\\
    \rowcolor{gray!25}5 & In the past 7 days, I felt like a failure. & In the past 7 days, how often have you felt like a failure? & How often do you feel like you're not making the most of your talents and abilities?; How often do you feel like you're not contributing enough to society?; How often do you feel like you've fallen short of your own expectations? & 8\\
    6 & In the past 7 days, I felt unhappy. & In the past 7 days, how often have you felt unhappy? & How often do you experience feelings of unhappiness?; Do you tend to dwell on negative thoughts and feelings?; Have you noticed yourself feeling unhappy more frequently than usual? & 7\\
    \rowcolor{gray!25}7 & In the past 7 days, I felt that I had nothing to look forward to. & In the past 7 days, how often have you felt that you had nothing to look forward to? & Do you frequently feel like your life lacks purpose or direction?; How often do you feel like there's nothing to look forward to in the coming days or weeks?; Have you been struggling to find joy in your daily activities? & 11\\
    8 & In the past 7 days, I felt that nothing could cheer me up. & In the past 7 days, how often have you felt that nothing could cheer you up? & Do you rarely feel happy or uplifted when you're feeling low?; Do you ever feel like you just can't shake off a negative mood?; Have you found it hard to see the positive side of things lately? & 11\\
    \bottomrule
  \end{tabular}
\end{table*}

\subsection{Generating Question Variants Using LLMs}
\label{sec:variants}

By using LLMs, we can yield a greater range of variations for each question faster and with fewer human resources, potentially increasing user engagement \cite{cox2023prompting}. However, it is important to note that such variations may include harmful language or language that deviates from the main concept of the original questionnaire items. Particularly in the domain of mental health, unconstrained outputs from LLMs may not be suitable for measuring various aspects of mental health issues, as LLMs are known to provide dangerous advice or misinformation \cite{olafsson2023accomodating,quach2020researchers,bickmore2018patient,miner2016smartphone}.

To address the potential issue of harmful outputs (hallucinations or misinformation) from LLMs, we used ChatGPT (March 2023 version) and GPT-3 to generate different variants of each question and manually filtered them before using them in our VA system (\autoref{fig:process}). 
For prompting, we provided the original item and response scale and asked the LLM to paraphrase the main item into a new question. These models were prompted to generate variants that fit the response scale. These variations often contained words, phrases, or concepts closely related to the latent construct we wanted to measure. A total of 178 unique variants of the eight questions were generated. A psychologist then ranked and filtered the variants to create a final list that matched the meaning and purpose of the original question. We also removed potentially dangerous questions that might involve suicide or thoughts about killing oneself. We obtained 67 variants of the questions, with a few samples available in \autoref{tab:variants}. All implementation details of generating the item variants and conversational small talk including jokes and empathetic responses and the full list of LLM-generated content used for the study can be found in supplementary materials\footnote{\url{\repourl}}.

\section{Longitudinal Evaluation Study}
\label{sec:evaluation_study}

From April to May 2023, we conducted a 15-day longitudinal online study to evaluate the psychometric properties of our questionnaire variants, user engagement, and user perception of the VA questionnaire system. For the first 14 days, participants were asked to talk with the VA once per day, which lasted a few minutes, and answer a short online survey after each interaction. On the 15\textsuperscript{th} day, participants completed a final online survey. \autoref{fig:diagram_teaser} provides the entire study workflow.

The experiment followed a randomized between-subject design with three study conditions.
In one condition --- CONTROL, we had the VA administer the standardized eight-item PROMIS\textsuperscript{\textregistered} depression questionnaire in question format for a more conversational experience. In the two intervention conditions --- ITEM VARIANTS ONLY and ITEM VARIANTS PLUS, the agent randomly chooses question variants described in \autoref{sec:variants}. In addition, ITEM VARIANTS PLUS includes additional social and conversational content for dialogue, such as anecdotes, jokes, empathetic responses, inspiring or hopeful messages, and farewells generated by LLMs (\autoref{tab:small_talk}). In a typical session for the ITEM VARIANTS PLUS condition, the agent first shared a short personal anecdote and then a randomly selected joke before administering the questionnaire. The agent provides empathetic responses based on user responses to the questions and ends with a randomly chosen motivational message and farewell statements.

\subsection{Measures \& Data Collection}
\label{sec:measures}

We collected the following four items with a 7-point scale response after each interaction with the agent: ``How satisfied are you with the agent?'' (1=``not at all'' and 7=``very satisfied''),  ``How much would you like to continue talking with the agent?'' (1=``not at all'' and 7=``very much''), ``How natural was your conversation with the agent?'' (1=``not at all'' and 7=``very natural''), and ``Did the agent feel repetitive?'' (1=``not at all'' and 7=``very repetitive'').
User engagement is assessed as the number of completed interactions with the agent.

After the two-week study period, we administered the final survey on the 15\textsuperscript{th} day, which included the eight-item Patient Health Questionnaire depression scale (PHQ-8) \cite{kroenke2009phq, razykov2012phq}, the system usability scale (SUS) \cite{bangor2008empirical}, overall system satisfaction measures, agent satisfaction measures, and measures related to user perception of questions asked by the agent (\autoref{tab:survey-responses}).
We also asked four open-ended questions about their experiences.

\subsection{Participants}
\label{sec:participants}

Participants were recruited via an online research platform (\url{www.prolific.com}). They were required to be 18 years old or older, able to read and write English, located in the USA, have working audio for their computer, and have a browser that supports WebGL 2.0. Participants were told to interact daily with the system at least seven times during the two weeks and complete a final survey for compensation. Each interaction consisted of a conversation with the VA and a short survey. The minimum interaction requirement was to ensure each participant was provided with the questionnaire several times.
Due to the sensitive nature of asking about depression symptoms, participants were told that the system is for assessment only and were provided with a list of mental health resources in the USA. The study was approved by our institutional review board.

\begin{table*}[t]
  \caption{Examples of Conversational Content Generated by LLMs.
  }
  \label{tab:small_talk}
  \small
  \begin{tabular}{m{.13\textwidth}|m{.76\textwidth}c}
    \toprule
    \textbf{Category} & \textbf{Example Content} & \textbf{Count}\\
    \midrule
    \rowcolor{gray!25}Personal Anecdotes &\tabitem I love going for hikes in the beautiful outdoors! This morning, I took a hike around a nearby lake. The fresh air and peaceful atmosphere made it the perfect way to start the day!\newline\tabitem I just finished reading this amazing book I stumbled upon! I couldn't put it down. It was a captivating journey that kept me on the edge of my seat and I can't wait to recommend it to all my friends. \newline\tabitem This past weekend I decided to try a new restaurant in town. The atmosphere was cozy and the food was delicious! I'm already looking forward to my next visit so I can try something else off the menu. & 37\\
    Jokes & \tabitem Why don't scientists trust atoms? Because they make up everything!\newline\tabitem Why did the smartphone need glasses? Because it lost all its contacts!\newline\tabitem What do you call a bear with no teeth? A gummy bear! & 24\\
    \rowcolor{gray!25}Empathetic\newline Responses & \tabitem I understand how overwhelming helplessness can be, and I'm here to support you.\newline\tabitem I'm sorry to hear that you feel this way. Please remember that you are valuable and that your feelings are valid. \newline\tabitem I understand how you're feeling. It's normal to feel overwhelmed at times and it's ok to take a step back and take care of yourself. & 35\\
    Inspiring or\newline Hopeful Messages & \tabitem You are not alone in your struggles. Reach out to others for support and comfort.\newline\tabitem Shiv Khera once said, Your positive action combined with positive thinking results in success.\newline\tabitem Today is your day to shine! Believe in yourself and make it happen. & 23\\
    \rowcolor{gray!25}Farewells or\newline Ending\newline Conversations & \tabitem Well, I should get going. It was nice talking to you!\newline\tabitem It was great catching up with you. I hope we can chat again soon!\newline\tabitem I enjoyed our conversation. It was nice talking with you. Have a great day! & 42\\
    \bottomrule
  \end{tabular}
\end{table*}

\section{Results}
\label{sec:results}

A total of 105 participants began the longitudinal evaluation study, with 35 participants assigned to each study condition. In total, 93 participants met the compensation requirements and completed the study successfully. All participants were on average 39 (\textit{SD}=12, \textit{Mdn}=37, \textit{Range}=21$\sim$73) years old. The gender breakdown was 49.5\% women, 46.7\% men, 2.9\% non-binary, and 1.0\% others. Participants were 75.2\% white, 7.6\% multiracial, 6.7\% Black/African-American, 3.8\% Asian/Asian American, 2.9\% Hispanic/Latinx, 1.9 \% American Indian/Alaska Native, and 1.9\% other. Participants had at least a high school degree or equivalent (43.8\% with a bachelor's degree, 24.8\% with some college, 10.5\% with a master's degree, 10.5\% with an associate degree, and 1.9\% with a doctoral/professional degree). When asked if they are currently in therapy or taking medication for depression, 80.0\% of participants said ``no'', 19.1\% said ``yes'', and 1.0\% preferred not to answer.

\subsection{Psychometric Properties of LLM-generated Item Variants}

We calculated Cronbach's alpha \cite{cronbach1951coefficient} to measure the internal consistency or reliability of the eight depression questions, or how closely related the eight questions are as a group. This involved looking at the responses of each participant for each administration. Cronbach’s alpha for the CONTROL condition was $\alpha$=.76 whereas the item variants were $\alpha$=.65. Although $\alpha$ for the version with variants was lower than that of the CONTROL, it showed acceptable internal consistency. \href{\repourl}{Supplementary materials} provide the percentages of responses for each question.

We also investigated whether the psychometric properties of items (questions) were consistent across the three groups. To investigate the validity of the PROMIS\textsuperscript{\textregistered} depression questionnaire across the three study conditions, we conducted a measurement alignment analysis using the \verb|sirt| package in \verb|R|, which allows the assessment of how the properties of individual items differ across groups \cite{han2024using}. In this model, each item is related to a single latent factor (depression) by a linear relationship described by a slope (i.e., factor loading) and intercept parameter. This analysis examined the degree to which groups can be ``aligned'' on the same scale. First, a confirmatory factor analysis (CFA) model allowed item slopes and intercepts to vary across groups. An $R^2$ was used to express how much variance in group differences was captured by true mean differences in the groups, rather than by different item properties across groups \cite{asparouhov2014auxiliary}. Our results, using the alignment procedure, indicated that 99\% of the between-group variation associated with slopes and 98\% of the between-group variation associated with intercepts could be attributed to factor mean and variance differences across the groups. Thus, the properties of the items were very consistent across groups. In a simulation, Asparouhov and Muthen \cite{asparouhov2014auxiliary} found that $R^2$ values of at least .98 were required to procure reliable factor rankings and that in general, $R^2$ values greater than .75 (i.e., up to 25\% non-invariance) were needed to produce trustworthy alignment results. Therefore, based on having achieved $R^2$ values greater than .98 for both the aligned item intercepts and loadings, we demonstrated the consistency of the PROMIS\textsuperscript{\textregistered} questionnaire administered across the three study conditions and concluded that only 2\% and 1\% of the variance could be attributed to differences in the item slopes and intercepts across the three study conditions.

To test the validity of each of the three administrations of the PROMIS\textsuperscript{\textregistered} questionnaire, we compared the three administrations against an external criterion---the PHQ-8. Specifically, the PHQ-8 was added to the CFA, mentioned above. We found that the correlations between the PROMIS\textsuperscript{\textregistered} questionnaire and the PHQ-8 were greater than or equal to .80 across all study conditions, demonstrating convergent validity of the LLM-generated items.

\subsection{Engagement}

We analyzed differences in the number of completed interactions by study conditions, including participants who did not complete the study.
Participants in the CONTROL group had an average of 9.9 (\textit{SD}=3.8) interactions with the agent while ITEM VARIANT ONLY and ITEM VARIANT PLUS groups had an average of 11.3 (\textit{SD}=3.0) and 10.8 (\textit{SD}=3.1) interactions, respectively, with no significant differences between conditions, (\textit{F}(2, 102)=1.81, \textit{p}=.17). 
Looking at the number of participants who met the minimum interaction requirement, we found a trending difference among the three groups (\textit{$X^{2}$}(2, \textit{N}=105)=5.1, \textit{p}=.08). CONTROL condition had 80\% of participants who met the requirement while ITEM VARIANTS ONLY condition had 97\% and ITEM VARIANTS PLUS condition had 89\%. We found no significant differences between participants who received treatment for depression and those who did not.

\subsection{Perception of System \& Agent}
\label{sec:perception}

\begin{table*}
\centering
\caption{User perceptions of the system, agent, and the questions. System-related items are on 7-point scales (from ``not at all'' to ``very much''), with all other items on 5-point  scales, with medians per group reported.}
\label{tab:survey-responses}
\vspace*{-\baselineskip} 
\small
\begin{tabular}{cm{7.5cm}ccc}\\
\toprule
\textbf{Category} &
  \textbf{Item} &
  \textbf{CONTROL} &
  \textbf{ITEM VARIANTS ONLY} &
  \textbf{ITEM VARIANTS PLUS} \\ 
  \midrule
\multirow{5}{*}{System} &
  Mean system usability scale (0-100) &
  $78.6 \pm 12.9$ &
  $75.2 \pm 14.8$ &
  $75.3 \pm 17.3$ \\ 
  \cmidrule(l){2-5} 
 &
  \cellcolor{gray!25}How satisfied are you with the system? &
  \cellcolor{gray!25}4.0 &
  \cellcolor{gray!25}4.5 &
  \cellcolor{gray!25}5.0 \\
 &
  How much would you like to continue using the system? &
  3.0 &
  4.0 &
  3.0 \\
 &
  \cellcolor{gray!25}Would you recommend the system to your friends and family? &
  \cellcolor{gray!25}4.0 &
  \cellcolor{gray!25}4.0 &
  \cellcolor{gray!25}3.0 \\
 &
  \cellcolor{azure!75}\textbf{Mean of composite score} &
  \cellcolor{azure!75}$3.6 \pm 1.7$ &
  \cellcolor{azure!75}$4.0 \pm 1.8$ &
  \cellcolor{azure!75}$3.9 \pm 1.9$ \\ 
  \midrule
\multirow{9}{*}{Agent} &
  How satisfied are you with the agent? &
  3.0 &
  4.0 &
  4.0 \\
 &
  \cellcolor{gray!25}How much would you like to continue talking with the agent? &
  \cellcolor{gray!25}3.0 &
  \cellcolor{gray!25}4.0 &
  \cellcolor{gray!25}3.0 \\
 &
  How much do you trust the agent? &
  3.0 &
  3.0 &
  3.0 \\
 &
  \cellcolor{gray!25}How much do you like the agent? &
  \cellcolor{gray!25}3.0 &
  \cellcolor{gray!25}4.0 &
  \cellcolor{gray!25}4.0 \\
 &
  How knowledgeable was the agent? &
  3.0 &
  3.0 &
  3.0 \\
 &
  \cellcolor{gray!25}How natural was your conversation with the agent? &
  \cellcolor{gray!25}2.0 &
  \cellcolor{gray!25}2.5 &
  \cellcolor{gray!25}2.0 \\
 &
  Did the agent feel repetitive?  &
  5.0 &
  4.0 &
  4.0 \\
 &
  \cellcolor{gray!25}How would you characterize your relationship with the agent? (complete stranger - close friend) &
  \cellcolor{gray!25}2.5 &
  \cellcolor{gray!25}3.0 &
  \cellcolor{gray!25}2.0 \\
 &
  \cellcolor{azure!75}\textbf{Mean of composite scores} &
  \cellcolor{azure!75}$3.0 \pm 0.85$ &
  \cellcolor{azure!75}$3.2 \pm 0.92$ &
  \cellcolor{azure!75}$3.1 \pm 1.03$ \\ 
  \midrule
\multicolumn{1}{l}{\multirow{5}{*}{Questions}} &
  How coherent were the questions asked by the agent? &
  4.0 &
  4.0 &
  4.0 \\
\multicolumn{1}{l}{} &
  \cellcolor{gray!25}How natural were the questions asked by the agent? &
  \cellcolor{gray!25}4.0 &
  \cellcolor{gray!25}3.0 &
  \cellcolor{gray!25}4.0 \\
\multicolumn{1}{l}{} &
  Were the questions asked by the agent easy to understand? &
  4.0 &
  4.5 &
  5.0 \\
\multicolumn{1}{l}{} & \cellcolor{gray!25}How often were the questions asked by the agent related to the topic of mental health? (never - almost constantly) & \cellcolor{gray!25}5.0 & \cellcolor{gray!25}5.0 & \cellcolor{gray!25}4.0 \\
\multicolumn{1}{l}{} &
  \cellcolor{azure!75}\textbf{Mean of composite score} &
  \cellcolor{azure!75}$4.2 \pm 0.57$ &
  \cellcolor{azure!75}$4.1 \pm 0.53$ &
  \cellcolor{azure!75}$4.1 \pm 0.68$ \\ 
  \bottomrule
\end{tabular}%
\end{table*}

At the end of the two-week study period, participants rated the overall system as usable with a mean SUS score of 76.3 (\textit{SD}=15.1).
They also reported an above neutral rating (\textit{Mdn}=4, \textit{IQR}=2) on a 7-point scale for overall satisfaction with the system.
Participants rated their satisfaction with the agent with a median of 3.5, which was significantly higher than a neutral score of 3, \textit{Z}=1.9, \textit{p}=.03, \textit{r}=.25. In addition, they reported the agent's repetitiveness at 4.5, significantly greater than a neutral score of 3, \textit{Z}=6.6, \textit{p}<.001, \textit{r}=.71. 
There were no significant differences among the three conditions for any of these measures (\autoref{tab:survey-responses}).
From the content analysis of open-ended responses,
we found that those in CONTROL were significantly more likely to mention ``repetitiveness'' compared to those in the two variant groups,
\textit{$X^{2}$}(1, \textit{N}=93)=5, \textit{p}=.029. 

Participants reported higher overall satisfaction with the agent's questions, based on the median composite scores (\textit{Mdn}=4.1) being greater than a neutral of 3, \textit{Z}=8.2, \textit{p}<.001, \textit{r}=.86. Across all conditions, participants reported responses significantly above neutral of 3 for coherence (\textit{Mdn} = 4.5, \textit{Z} = 7, \textit{p}<.001, \textit{r}=.86), naturalness (\textit{Mdn}=4, \textit{Z}=4.1, \textit{p}<.001, \textit{r}=.43), how easy the questions were to understand (\textit{Mdn}=4.5, \textit{Z}=8.2, \textit{p}<.001, \textit{r}=.87), and relevance (\textit{Mdn}=4.5, \textit{Z}=8.5, \textit{p}<.001, \textit{r}=.88). No significant differences among study conditions were found (\autoref{tab:survey-responses}).

For the repeated measures collected after each interaction with the agent, we did not find any significant differences across the study conditions.
Although not significantly different, participants in the CONTROL group reported a mean score of 5.1 (\textit{SD}=1.3) for agent repetitiveness over 14 days while ITEM VARIANTS ONLY and ITEM VARIANTS PLUS conditions had means of 4.9 (\textit{SD}=1.4) and 4.7 (\textit{SD}=1.4), respectively. 

\subsection{Qualitative Results}
We conducted a deductive thematic analysis of the open-ended responses (3,003 words), guided by sensitizing concepts that focused on participant satisfaction and feedback on additional features \cite{clarke2015thematic}.
We used elements of the grounded theory method, including open, axial, and selective coding \cite{corbin1990grounded}.

\paragraph{\textbf{Comforting vs Uncanny Agents.}} Some participants expressed positive sentiments about talking to the agent and mentioned their willingness to interact daily: \textit{``I like how someone was checking in with me daily to make sure I was alright.''} [P43 - ITEM VARIANTS PLUS] and \textit{``I liked the character, she felt like a safe person to talk to.''} [P67 - ITEM VARIANTS ONLY]. One participant mentioned that their least favorite part of the system was that they were not able to have more interactions with the agent, \textit{``I can't really give the answers I want or talk with her as long as I want''} [P13 - ITEM VARIANTS PLUS]. Another participant mentioned their desire to have deeper interaction with the agent on sharing their feelings, \textit{``Maybe an option to expand on questions if I'm feeling down, like a deeper dive into my feelings, but still utilizing the multiple-choice selections''} [P3 - CONTROL]. Conversely, some found the interaction with the VA to be uncanny and unnatural. For instance, P80 [CONTROL] found the interaction with the agent strange, \textit{``The attempt to make the robot AI feel human looking---it was uncanny valley to the max.''} 

\paragraph{\textbf{Various Reasons for Repetitiveness.}} 
Most participants, especially in the CONTROL group, mentioned the repetitiveness of the system and agent. P82 [CONTROL] said, \textit{``The repetition, being asked the same questions every single day, was a chore even though it wasn't very difficult. It lost its charm after the first few days.''} P89 [CONTROL] also commented on the repetitiveness of questions, \textit{``same questions over and over''}. Some participants, across all conditions, talked about how the user response options were repetitive. P88 [CONTROL] expressed that their least favorite part of the system was \textit{``How repetitive the responses were''}. Others, even in the intervention groups, expressed how the agent's responses felt repetitive. For instance, P66 [ITEM VARIANTS ONLY] said, \textit{``the feedback was repetitive''}. 

\paragraph{\textbf{Humor and Small Talk Does Not Always Work.}} Some participants mentioned \textit{``hearing the jokes she had''} [P85] as their favorite part of the system, while others said that they would like to skip \textit{``the bad dad jokes''} [P11]. Furthermore, P36 found the anecdotes and jokes to be forced, saying that they would like \textit{``No forced stories and jokes in the beginning of the session.''} P87 did not like the agent telling stories from her daily life saying, \textit{``Probably the `let me tell you about myself' stupidity. It was ridiculously patronizing that I was expected to take that seriously. A toddler would know an AI isn't getting sore throats and going to the movies''}. While some participants appreciated the humor and small talk in their interaction with the agent, others felt that the VA's small talk detracted from their ability to focus on answering the questionnaire.

\section{Discussion}
\label{sec:discussion}

We demonstrated the reliability and validity of LLM-generated question variants in a two-week validation study. The measurement alignment analysis and Cronbach's alpha showed the reliability of the questions administered in all three study conditions. In addition, the three different administrations of the PROMIS\textsuperscript{\textregistered} depression questionnaires demonstrated good validity when compared to an external criterion of PHQ-8 which is another validated, standardized questionnaire for screening depression. These findings support \textbf{H1} by showing that the LLM-generated questionnaire variants do retain reliability and validity when compared to an external criterion. Furthermore, participants given the item variants found the questions coherent, natural, easy to understand, and relevant to the conversational topic based on the above neutral median self-reported data.

In total, 105 participants started the study and 93 participants met the minimum interaction requirement. The CONTROL group had the lowest percentage of participants (80\%) who met the minimum interaction requirement. We saw a trending difference in the number of participants who met the minimum interaction requirement of seven interactions among the three study conditions. In addition, participants in the CONTROL group found the agent's questions more repetitive compared to participants with the LLM-generated questionnaire variants based on our content analysis reported in \autoref{sec:perception}. However, we did not find any other significant results from the post-study survey results. These findings partially support \textbf{H2} as we observed that questionnaires delivered in a different form daily did show trending differences in the number of participants who met the minimum interaction requirement.

Furthermore, we did not find any differences in engagement, satisfaction, or usability between ITEM VARIANTS ONLY and ITEM VARIANTS PLUS conditions. Therefore, our findings do not support \textbf{H3}, in which questionnaires delivered with humor and small talk will increase engagement compared to those without them in an interview with a VA. Qualitative findings showed that some participants found the jokes and stories entertaining and interesting while others found them to be forced. This demonstrates how simple jokes and small talk might not always be a reliable mechanism to increase engagement or satisfaction for repeated interviews. 

\subsection{Limitations \& Future Work}
\label{sec:limitations_future_work}
There are several limitations to our study beyond the small convenience sample used. We conducted our evaluation study using only one standardized questionnaire, so it is unclear whether our results hold for other questionnaires. Our compensation structure with a strict minimum interaction requirement may have affected the results of engagement and interaction with the agent and could be seen as a limitation of our study design. 
In addition, the technical limitations of the prototype could have affected participant satisfaction.

Future work should consider ways of adding more variations in the dialogue structure, VA responses to users, and user question response options (scale anchor response options) to further reduce repetitiveness. Future work could also study the effects of letting participants interact with the agent using an unstructured input. This approach could further reduce perceptions of repetitiveness. However, finding a balance between personalization and standardization would need further examination. In addition, varying questions could increase the cognitive effort to process and respond. Future studies to understand the trade-offs between respondent fatigue/cognitive load and variations are needed.

Further research on incorporating humor and anecdotes generated by LLMs for longitudinal VA research should be considered. Our LLM-generated jokes were similar to how \citeauthor{jentzsch2023chatgpt} \cite{jentzsch2023chatgpt} found ChatGPT to only produce limited joke patterns. Having more diverse jokes and stories (backstories of the agent or even stories of real people) and adapting to user conversation responses can be interesting future directions.

A previous study \cite{jaiswal2019virtual} has compared using VA and form-based questionnaires, and for our future studies, we will compare the effect of VA-based interactions with item variants with form-based questionnaires. Furthermore, future studies could examine utilizing more novel approaches, such as logical control \cite{basar2023hyleca}, chain-of-thought prompting \cite{wei2022chain}, or augmentation to use external tools \cite{mialon2023augmented}, to create a safeguard for utilizing LLMs more directly to conversational agents. 

\section{Conclusion}
\label{sec:conclusion}

We demonstrated that LLM-generated item variants for a depression questionnaire maintain good psychometric properties when delivered by a virtual agent. The LLM-generated item variants demonstrated validity and reliability and were seen to be coherent, natural, easy to understand, and relevant to the topic at hand. Additionally, participants who received these LLM-generated item variants generally found the agent less repetitive over a two-week study period compared to the CONTROL group. However, we found that including conversational humor and small talk in questionnaire administration interviews by an agent did not result in higher satisfaction or engagement.
Striking a balance between personalization and standardization will be crucial for maintaining high-quality data collection and boosting response rates in delivering longitudinal self-report questionnaires. While using LLMs for producing questionnaire variants necessitates meticulous prompt preparation and manual output review, it offers the advantage of efficiently scaling and expediting the generation of diverse content. We view this study as a step forward in integrating LLMs into VAs to diversify and enhance questionnaire administration while maintaining validity and reliability.

\begin{acks}
Research reported in this publication was supported by the National Institute of Cancer of the National Institutes of Health under award number R01CA271145.	
\end{acks}

\bibliographystyle{ACM-Reference-Format}
\bibliography{bibliography}


\end{document}